\documentclass[sn-mathphys,Numbered, iicol]{sn-jnl}

\usepackage{graphicx}%
\usepackage{multirow}%
\usepackage{amsmath,amssymb,amsfonts}%
\usepackage{amsthm}%
\usepackage{mathrsfs}%
\usepackage[title]{appendix}%
\usepackage{xcolor}%
\usepackage{textcomp}%
\usepackage{manyfoot}%
\usepackage{booktabs}%
\usepackage{algorithm}%
\usepackage{algorithmicx}%
\usepackage{algpseudocode}%
\usepackage{listings}%
\usepackage{pgfplots}
\usepackage{threeparttable}
\usepackage{subcaption}
\usepackage{enumitem}
\usepackage{tikz}
\usetikzlibrary{math,calc,positioning,shapes,decorations.pathreplacing,shapes.multipart,topaths,matrix,fit}
\pgfplotsset{compat=newest}
\tikzset{
    split/.style={rectangle split, rectangle split parts=#1,rectangle split horizontal, rectangle split part align=base, draw},
    vsplit/.style={rectangle split, rectangle split parts=#1, draw, inner sep=1.5pt},
    square/.style={minimum width=0.64cm, minimum height=0.64cm, draw, inner sep=0pt},
    squarev/.style={minimum width=#1 cm, minimum height=#1 cm, draw, inner sep=0pt}
}

\lstdefinestyle{FortranStyle}{
  language=Fortran,
  basicstyle=\ttfamily,
  keywordstyle=\color{blue},
  commentstyle=\color{green!40!black},
  stringstyle=\color{red},
  numbers=left,
  numberstyle=\tiny\color{gray},
  frame=single,
  breaklines=true,
  breakatwhitespace=true,
  tabsize=4
}
\theoremstyle{thmstyleone}%
\theoremstyle{thmstyletwo}

\theoremstyle{thmstylethree}%

\begin{document}

\title{O2ATH: An OpenMP Offloading Toolkit for the Sunway Heterogeneous Manycore Platform}


\author[1]{\fnm{Haoran} \sur{Lin}}\email{haoran.lin@mail.sdu.edu.cn}
\author[1]{\fnm{Lifeng} \sur{Yan}}\email{lifeng.yan@mail.sdu.edu.cn}
\author[1]{\fnm{Qixin} \sur{Chang}}\email{cqx@mail.sdu.edu.cn}
\author[3]{\fnm{Haitian} \sur{Lu}}
\author[3]{\fnm{Chenlin} \sur{Li}}
\author[3]{\fnm{Quanjie} \sur{He}}
\author[3]{\fnm{Zeyu} \sur{Song}}
\author*[1,3]{\fnm{Xiaohui} \sur{Duan}}\email{sunrise.duan@sdu.edu.cn}
\author[1]{\fnm{Zekun} \sur{Yin}}
\author[2]{\fnm{Yuxuan} \sur{Li}}
\author[2,3]{\fnm{Zhao} \sur{Liu}}
\author[2,3]{\fnm{Wei} \sur{Xue}}
\author[2,3]{\fnm{Haohuan} \sur{Fu}}
\author[2,3]{\fnm{Lin} \sur{Gan}}
\author[2,3]{\fnm{Guangwen} \sur{Yang}}
\author[1,3]{\fnm{Weiguo} \sur{Liu}}

\affil[1]{\orgdiv{School of Software}, \orgname{Shandong University}, \orgaddress{\city{Jinan}, \postcode{250100}, \country{China}}}

\affil[2]{\orgname{Tsinghua University}, \orgaddress{\country{China}}}

\affil[3]{\orgname{National Supercomputing Center in Wuxi}, \orgaddress{ \country{China}}}


\abstract{
The next generation Sunway supercomputer employs the SW26010pro processor, which features a specialized on-chip heterogeneous architecture. Applications with significant hotspots can benefit from the great computation capacity improvement of Sunway many-core architectures by carefully making intensive manual many-core parallelization efforts. However, some legacy projects with large codebases, such as CESM, ROMS and WRF, contain numerous lines of code and do not have significant hotspots. The cost of manually porting such applications to the Sunway architecture is almost unaffordable. To overcome such a challenge, we have developed a toolkit named O2ATH. O2ATH forwards GNU OpenMP runtime library calls to Sunway's Athread library, which greatly simplifies the parallelization work on the Sunway architecture.O2ATH enables users to write both MPE and CPE code in a single file, and parallelization can be achieved by utilizing OpenMP directives and attributes.
In practice, O2ATH has helped us to port two large projects, CESM and ROMS, to the CPEs of the next generation Sunway supercomputers via the OpenMP offload method. In the experiments, kernel speedups range from 3 to 15 times, resulting in 3 to 6 times whole application speedups.Furthermore, O2ATH requires significantly fewer code modifications compared to manually crafting CPE functions.This indicates that O2ATH can greatly enhance development efficiency when porting or optimizing large software projects on Sunway supercomputers.
}




\maketitle

\section{Introduction}\label{sec1}
\begin{figure*}
    \centering
    \includegraphics[width=1\textwidth]{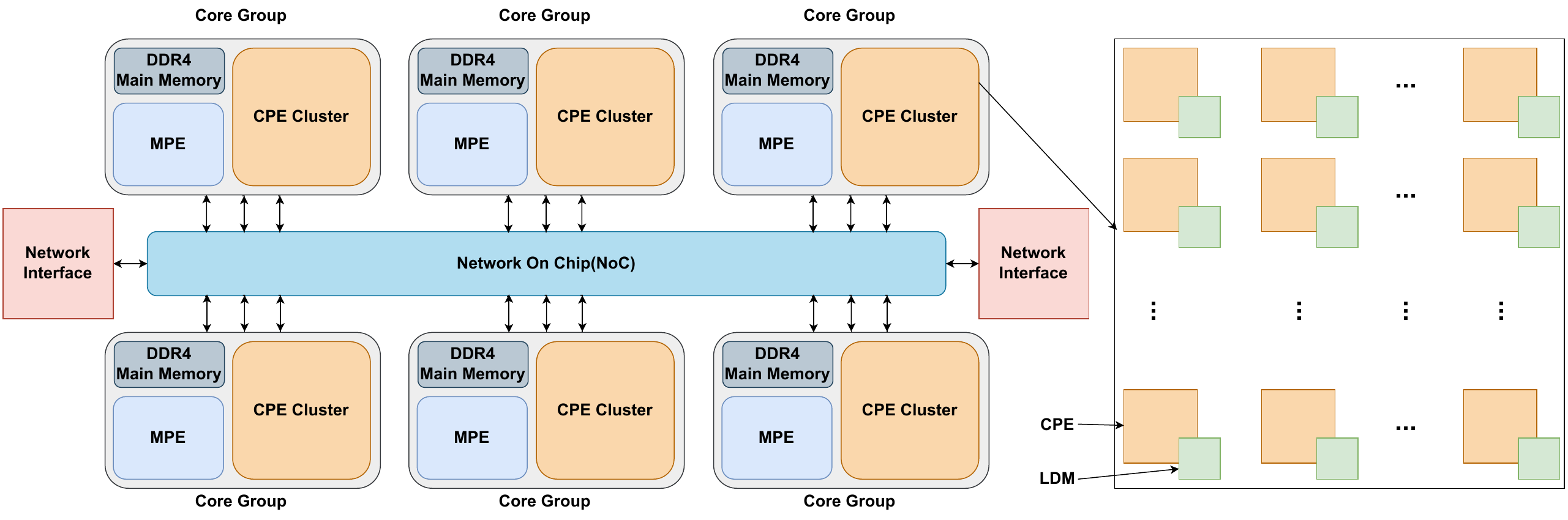}
    \caption{Architecture of the SW2610pro processor.}
    \label{fig:architecture}
\end{figure*}

The Sunway supercomputer boasts exceptional capabilities in the realm of high-performance computing with its ShenWei many-core series processors.  
The ShenWei many-core series processors feature distinctive on-chip heterogeneous architecture, currently containing the SW26010 and SW26010pro. 
Nowadays, the Sunway TaihuLight supercomputer \cite{fu2016sunway} still ranks within the top 10 of the world supercomputer list \cite{hpc-top500}. The SW26010 processor is adopted in Sunway TaihuLight as its exclusive computing power source.
Moreover, the next generation Sunway supercomputer boasts even better performance.
The immense computing power of the next generation Sunway supercomputers is derived from the new SW26010pro processor, and Figure \ref{fig:architecture} illustrates the architecture of the SW26010pro processor.


Each processor comprises six \textit{core groups} (CGs), and \textit{core groups} are interconnected by a ring network on chip. Within each CG, there is one \textit{management process element} (MPE) and 64 \textit{computing process elements} (CPEs). Thus, one SW26010pro processor comprises 390 cores \cite{liu2021redefining}. Raw serial code runs on MPE, and the high ratio of CPE count to MPE count causes this architecture to often suffer from Amdahl's Law \cite{amdahl} when porting large applications. Amdahl’s law says that speedup is given by 
\begin{equation}
Speedup = \frac{1}{s+\frac{p}{N}}
\end{equation}\label{eq:amdahl}
In the equation \ref{eq:amdahl} \cite{gustafson1988reevaluating}, $N$ is the number of processors, $s$ represents the time consumed by a single processor when executing the serial parts of a program, and the amount of time spent (by a serial processor) on parts of the program that can be done in parallel is represented by $p$. As for the many-core acceleration of applications, it can be rewritten as follows:
\begin{equation}\label{eq:parallel}
    Speedup=\frac{1}{1-\sum_{i=0}^N {p_i}+\sum_{i=0}^N\frac{p_i}{Speedup_i}}
\end{equation}

\begin{figure}[tb]
    \centering
    \begin{tikzpicture}
    \begin{axis}[height=6cm, width=\linewidth, cycle list name=exotic, ylabel={Overall speedup}, xlabel={Average kernel speedup}, legend style={at={(0.5,1)}, anchor=south, draw=none, font=\scriptsize}, legend columns=3]
      \addplot+[domain=0:30, samples=10] {1/(0.5/x+0.5)};
      \addplot+[domain=0:30, samples=10] {1/(0.9/x+0.1)};
      \addplot+[domain=0:30, samples=10] {x};
      \legend{$\sum_{i=0}^N{p_i}=0.5$,$\sum_{i=0}^N{p_i}=0.9$, $\sum_{i=0}^N{p_i}=1$}
    \end{axis}
  \end{tikzpicture}
    \caption{A plot for equation (\ref{eq:parallel}) assuming that kernels have the same speedup.}\label{fig:parallel}
    \label{fig:enter-label}
\end{figure}
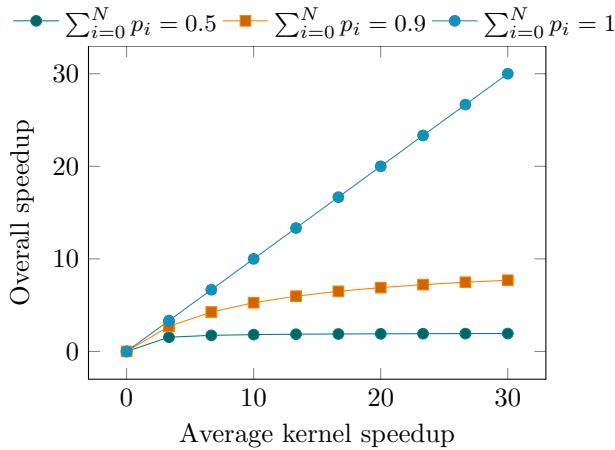

Here $N$ is the number of parallelized kernels, $p_i$ is the portion in serial execution time of $i$-th kernel. We can find that when speedup of each kernel is large, the sum of $p_i$ will limit the overall speedup. A plot of such condition assuming that kernels have the same speedup is shown in Figure \ref{fig:parallel}. It can be seen that when application is large, there should be a simple and quick method to parallelize kernels to increase the sum of $p_i$.

Each CPE is equipped 256KB user-controlled scratchpad memory named Local Data Memory (LDM), which can perform data transfers with the main memory using DMA and engage in communication with other LDMs through Remote Memory Access (RMA). 
Notably, the LDM is user-controllable, and a part of the LDM can be configured as a data cache. This has eased people from writing intensive DMA instructions like Sunway TaihuLight to make use of CPEs. The rest problem in rapid CPE parallelization is a more user-friendly threading framework.
Many projects have achieved significant performance improvements by migrating to and optimizing for the Sunway supercomputers (\cite{duan2018redesigning}, \cite{xu2022redesigning}, \cite{gao2020millimeter}, \cite{duan2023bio}).

Numerical computing is an approach for solving complex mathematical problems using only simple arithmetic operations \cite{bratu1914equations}, and it has significant applications in fields such as meteorology, geology, and materials science (\cite{danabasoglu2020community}, \cite{stolarski2018engineering}, \cite{grindon2004large}).
Nowadays, many numerical computing software have been ported to Sunway supercomputers in seek of performance improvements (\cite{gao2021lmff}, \cite{zhang2019sw_gromacs}, \cite{fu2017redesigning}).

However, to make the program utilize CPEs, users typically have to write and compile code separately for MPE and CPE. With the recent release of \textit{SWGCC}, users can also write hybrid parallel source files using emerging compiler plugins like SWUC \cite{cao2022design}. Hybrid programming can significantly simplify the CPE parallelization of C++ problem by utilizing the automatic capture generation feature of C++ lambda expressions, which can help users package data needed by CPE functions into a closure and pass them by the closure's address. But legacy projects written in Fortran or C cannot benefit from SWUC so much due to the lack of automatic capturing functionality.

Fortran is one of the earliest high-level programming languages \cite{backus1978history} and is widely used in the field of scientific and engineering computations.
Even today, numerous climate and weather applications such as CESM (Community Earth System Model) \cite{hurrell2013community} and ROMS (Regional Ocean Modeling System) \cite{shchepetkin2005regional} continue to utilize Fortran for numerical computations to solve complex problems. Furthermore, these applications have not been ported into newer computing languages mainly due to their huge codebases, so they require countless man-hours for CPE parallelization in the traditional way.

Traditional optimization workflow on Sunway architecture requires the user to write a function as CPE entrance. Then, the entrance function is dispatched to CPEs with the \texttt{athread\_spawn} interface in Sunway's Athread library. To pass multiple arguments to the CPE entrance, developers need to package them into a structure and pass its address to \texttt{athread\_spawn}. Declaring and packaging such arguments structure requires large repetitive and boring efforts. Also, a separated CPE entrance brings interruption in the program logic, making the optimized code difficult to read and maintain. To make things worse, algorithm developers of such software are mainly field experts, and they can hardly do modifications on parallelized CPE code. Each time a software is upgraded, developers need to rewrite a large portion of CPE codes.

To address the mentioned issues, some many-core architectures support OpenACC \cite{openacc2021} or OpenMP offloading \cite{chandra2001parallel} to parallelize code with minimum modifications by utilizing compiler-aided closure generation. SWACC, a ROSE-compiler-based source-to-source translator, can rewrite sources with OpenACC directives into separate MPE and CPE sources and dispatch CPE tasks with the Athread library. However, ROSE \cite{quinlan2000rose} uses a Java-based third-party Fortran parser, making the compilation very slow and memory-consuming. For example, it takes almost an hour to parse and rewrite a source file with thousands of lines of code. This limits the application of SWACC in large code bases.

With the release of the GCC-compatible Sunway compiler, SWGCC, there is a chance to modify intermediate representation (IR) with compiler plugins to achieve parallelization with much lower cost. 
GCC \cite{naishlos2004autovectorization} has a component named GNU OpenMP (GOMP) to support OpenMP, including offloading.
In the GOMP workflow, GCC parses the \texttt{target} directive, generates the separated target entrance, and calls to the GOMP runtime library, \textit{libgomp} \cite{libgomp}, to dispatch the target device task. While the offloading to CPEs is not supported by SWGCC, it can at least generate the target entrance and GOMP library calls correctly.
So, we can bridge the GOMP and Athread library via compiler plugins and proxy libraries. As a result, we implement O2ATH toolkit as a proxy between GOMP and Athread library.

O2ATH not only forwards GOMP library calls to Sunway's Athread library, but also has the following features:
\begin{enumerate}
    \item A flexible task server that dispatches computation tasks efficiently.
    \item Devirtualization of virtual function calls in CPE code.
    \item Inline trampolines in CPE code to avoid ``static chain" calls.
    \item Implementation of supporting functions for OpenMP \texttt{barrier}, \texttt{critical}, \texttt{single} and \texttt{master} directives.
\end{enumerate}
O2ATH is available at \url{https://gitee.com/swmore/swgomp}.

So that multi-core OpenMP code can be easily converted to Sunway many-core OpenMP code with minimal efforts.
With O2ATH, we can generate the code for CPE and MPE in a much simpler way handwriting the code. 

We have successfully accelerated two large projects, CESM and ROMS, on the next generation Sunway supercomputers using O2ATH.
We observed speedups ranging from 3 to 15 times across different kernels, resulting in 3 to 6 times overall performance improvements.
Additionally, we found that the utilization of O2ATH greatly simplifies the process of migrating applications to the Sunway supercomputer, resulting in significant human time savings and comparable performance with manually written code.

The rest of this paper is organized as follows. 
Section 2 describes the motivation behind the development of O2ATH. 
Section 3 provides a detailed description of the design of O2ATH. 
Instructions for using O2ATH are given in Section 4.
In Section 5, the benefits of O2ATH are evaluated. 
We discuss the future work in Section 6.
Finally, we conclude the paper in Section 7.

\section{Motivation}\label{sec2}
\subsection{Migration of Existing Application}\label{subsec2}

\begin{figure*}[ht]
  \centering
  \begin{subfigure}[b]{1\linewidth}
    \centering
    \begin{minipage}[t]{\linewidth}
      \begin{lstlisting}[style=FortranStyle]
subroutine exam(B)
  ...
end subroutine exam
...
DO j=js, je
    DO i=1, 10
        call exam(B(i,j))
        A(i,j)=A(i,j) - B(i,j)
    END DO
END DO
...
      \end{lstlisting}
    \end{minipage}
  \end{subfigure}%
  \caption{The original example code for parallelizable loops.}
  \label{origin-code}
\end{figure*}

\begin{figure*}[ht]
  \centering
  \begin{subfigure}[b]{1\linewidth}
    \centering
    \begin{minipage}[t]{\linewidth}
      \begin{lstlisting}[style=FortranStyle,label=list:mpe-cod]
...
!Predefined the roms_cpe_param structure
type(roms_cpe_param) :: mpe_param 
external :: slave_example_loop_cpe
...
mpe_param%js  =  js
mpe_param%je  =  je
mpe_param%A   => A
mpe_param%B   => B
call athread_spawn(slave_example_loop_cpe,mpe_param)
call athread_join()
      \end{lstlisting}
      \caption{Accelerated MPE code}
      \label{list:mpe-code}
    \end{minipage}
  \end{subfigure}%
  \hfill
  \begin{subfigure}[b]{1\linewidth}
    \centering
    \begin{minipage}[t]{\linewidth}
      \begin{lstlisting}[style=FortranStyle,label=list:cpe-cod]
subroutine exam(B)      !copied to CPE code
  ...
end subroutine exam
subroutine example_loop_cpe(mpe_param)
  ...
  type(roms_cpe_param) :: mpe_param, cpe_param
  integer :: js, je, i, j, cpe_tid
  real(r8), dimension (:,:)   ,pointer :: A, B
  call crts_dma_get(...)
  a  =  cpe_param%a
  ...
  B      => cpe_param%B
  DO j=cpe_tid+js, je, 64
    DO i=1, 10
      call exam(B(i,j))
      A(i,j)=A(i,j) - B(i,j)
    END DO
  END DO
end subroutine example_loop_cpe
      \end{lstlisting}
      \caption{Accelerated CPE code}
      \label{list:cpe-code}
    \end{minipage}
  \end{subfigure}
  \caption{After accelerating the parallelizable loop with a exam function on SW26010pro.}
  \label{fig:accelerated-code}
\end{figure*}

To migrate and optimize an existing Fortran code on the Sunway supercomputer, the initial step typically involves identifying parallelizable loops.
As shown in Figure \ref{origin-code}, assuming this is a loop that consumes a significant portion of time, we need to parallelize it to utilize CPEs in the SW26010pro processor.

This implies that we need to write code separately for MPE and CPE, as shown in Figure \ref{fig:accelerated-code}. We can see that Figure \ref{list:mpe-code} represents the MPE code, while Figure \ref{list:cpe-code} means the CPE code.
To perform such optimization, we need to identify data used by CPE code and declare a parameter structure to adopt the data, such as \texttt{roms\_cpe\_param} in Figure \ref{fig:accelerated-code}. Then, we need to wrap data into the structure in MPE code. Also, we need to recursively identify functions called in CPE code like the ``\texttt{exam}'' function here and duplicate such functions into the CPE source file.
Even if such ``\texttt{exam}'' functions are simple and thread-safe, they should be duplicated into the CPE source file to ensure they are compiled for CPEs.
Then, the MPE code needs to call \texttt{athread\_spawn} to invoke the task on CPEs.
We can imagine that when dealing with a complex application, especially with multiple levels of function calls and tens of parameters, the code involved can become quite intricate. 
What's more, especially for projects with substantial code bases, the task of code maintenance becomes even more challenging. Also, the robustness of such code is not well due to people frequently making mistakes during refactoring, and such bugs are not easy to identify because of the large code portion.

\subsection{Design Objectives of O2ATH}\label{subsec2}
Based on the issues mentioned above, we propose the following design goals:
\begin{itemize}
    \item A unified programming framework supporting C, C++ and Fortran that can eliminate the need for manual coding of MPE and CPE separately. 
    \item Modifications made to the original code can be automatically compiled into both MPE code and CPE code;
    \item Automation of the argument passing workflow, so that there is no need to construct parameter structures;
    \item A straightforward programming style. We need to write code in a simple, easily understandable way for rapid iterative development.
\end{itemize}

\section{Toolkit Design}\label{sec3}
We analyzed the SWGCC for the SW26010pro processor and found that this compiler  is compatible with GCC plugins, including partial support for GOMP. That is, SWGCC can generate interfacing IR to \textit{libgomp}.
Based on the interfacing mechanism, we developed O2ATH to forward GOMP library calls to Sunway's Athread Library.
\subsection{The Mechanism of GOMP}\label{subsec3}
Open Multi-Processing (OpenMP) is a widely used programming standard for multi-core parallel processing in C, C++ and Fortran programs.
OpenMP provides directives and compiler support to distribute workloads among multiple threads efficiently.
OpenMP also has a \texttt{target} directive to offload execution from the host to the target device(s) \cite{OpeMP}, and it supports the \textit{device} clause that specifies the target device ID and the \textit{private} clause that speciﬁes variables need to be privatized. Both the separated host and device memory model and the unified host and device memory model are supported by OpenMP offloading. We choose the unified one due to CPEs can access main memory directly, and the size of LDM (256KiB) does not match the typical size (several Gigabytes) of device memory.

\textit{libgomp}, short for GNU Offloading and Multi Processing Runtime Library, is the runtime library for GOMP and GNU OpenACC (GOACC) to support parallel programming and offloading computing tasks within the GNU Compiler Collection (GCC). 
Thus, \textit{libgomp} can perform various actions, such as creating and managing threads, recognizing data to ensure data consistency and synchronization among threads, executing tasks in parallel, and offloading tasks to specific target devices.
It empowers developers to offload computationally intensive tasks to accelerators like GPUs, unlocking the potential of heterogeneous computing environments for enhanced performance.

During the compilation process, GOMP can recognize OpenMP directives and modify IR for the parallelization. 
It separates host and device code into functions and captures variables used in the parallel region. 
Afterward, it replaces the parallelized code with function call to \texttt{\_\_builtin\_GOMP\_target\_ext}, which spawns the device function.
To adapt OpenMP offload code to the SW26010pro architecture, we can design plugins to leverage this interface mechanism. Based on the IR generated by GOMP, we need to forward GOMP library calls to Sunway's Athread library.
We accomplish the task with the O2ATH toolkit, which consists of a compiler plugin and a runtime library.

\subsection{Compiler Plugin}\label{subsubsec3}

\begin{figure*}[tb]
    \includegraphics[width=.95\linewidth]{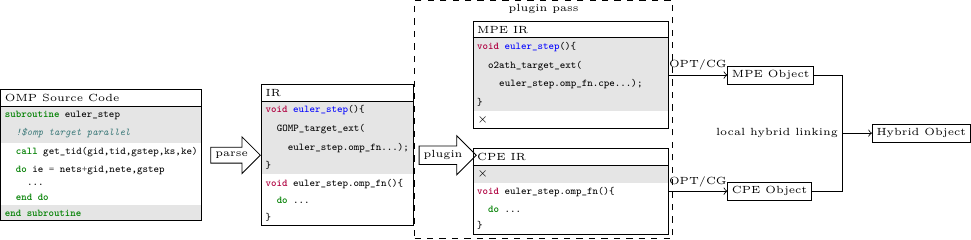}
    \caption{A detailed example for the compiling workflow when using the O2ATH plugin.}\label{fig:o2ath-plugin}
\end{figure*}
As shown in Figure \ref{fig:o2ath-plugin}, GOMP generates the IR for OpenMP parallelization, and the compiler plugin modifies the IR for MPE and CPE separately. The MPE and CPE objects are linked together as the compilation output of the plugin.

\subsubsection{Support for Offloading}

\begin{figure*}
    \centering
    \includegraphics[width=1\textwidth]{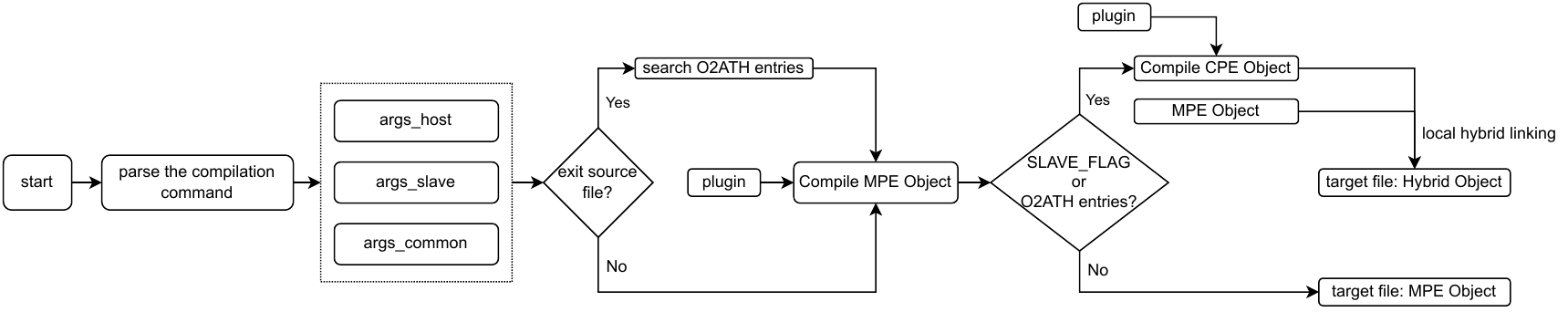}
    \caption{The workflow of xfort when using the plugin to compile source files.}
    \label{fig:xfort}
\end{figure*}

Because switchable target code generation is not supported by SWGCC, we must compile the source twice with different actions to obtain both MPE code and CPE code. The plugin can identify CPE entrance functions according to attributes added by GOMP or developers, and then the dependent functions are recursively marked for CPE code generation by the O2ATH plugin. When compiling MPE code, CPE entrance functions are removed, and vice versa. Also, the plugin replaces \texttt{GOMP\_target\_ext} with \texttt{o2ath\_target\_ext} in MPE code to bridge successive actions to the O2ATH runtime library. When compiling CPE code, the plugin has further actions to assist people in porting large code segments, which will be described in the following sections. To make the plugin easier to use, we have a wrapper script, \textit{xfort}, for compiling Fortran programs with the O2ATH plugin. The workflow of xfort is shown in Figure \ref{fig:xfort}:
    \begin{enumerate}
    \item \textit{xfort.py} parses the compilation commands and  separates different compilation parameters into \textit{args\_host}, \textit{args\_slave}, and \textit{args\_common}.
    \item Then, it tries to search \texttt{O2ATH entries} in the source file, which is an O2ATH extension to specify CPE functions for cross file function calls alternatively.
    \item \textit{xfort.py} compiles the MPE object with the compiler plugin enabled.
    \item During the compilation process, the plugin traverses the call graph to do further actions.
        \begin{itemize}
            \item It checks whether the source file contains functions with \texttt{omp declare target} or \texttt{omp target entrypoint} attribute to determine whether CPE code generation is required. Functions specified in \texttt{O2ATH entries} will be also marked as \texttt{omp declare target}.
            \item  If CPE code generation is required, the plugin creates an empty file whose name is specified environment variable \texttt{SLAVE\_FLAG}.
        \item Search for the references to \texttt{\_\_builtin\_GOMP\_target\_ext} and redirect them to \texttt{o2ath\_target\_ext}.
        \end{itemize}
    \item If there is \texttt{SLAVE\_FLAG} file, \textit{xfort} invokes the compiler again for CPE code generation. Finally, the MPE and CPE objects are locally linked together as the compilation output of the plugin. Otherwise, \textit{xfort} will regard the MPE object as the target file.
    \end{enumerate}

\subsubsection{Inlining Trampolines on CPEs}
Trampolines are functions nested in another function, which are widely used in Fortran programs. Code in a trampoline can access variables in the outside function via the ``static chain" calling convention of the platform's abstract binary interface (ABI). In Sunway ABI, ``static chain" call requires to execute instructions stored in the stack, which is not supported by CPE due to the limitation of 24-bit short PC. The O2ATH plugin can forcefully inline non-recursive trampolines, so that it can support trampolines in most conditions.

\subsubsection{Devirtualization of Fortran Class Functions}
All class functions in Fortran are virtual, while most class functions in CESM and ROMS do not require dynamic binding. GOMP captures objects with their virtual table, but function pointers in the virtual table still point to MPE code. Thus, when a virtual function is called on CPE, the CPE jumps to the MPE function according to the virtual table, causing unexpected behaviors. To reduce the manual interception in such cases, the O2ATH plugin can also aggressively convert such function calls to static binding. This can find CPE functions correctly in most cases, and the plugin also leaves a warning message for users that claims the call is de-virtualized aggressively.

\subsection{Runtime Library}\label{subsubsec3}

\begin{figure*}[tb]
    \includegraphics[width=.95\linewidth]{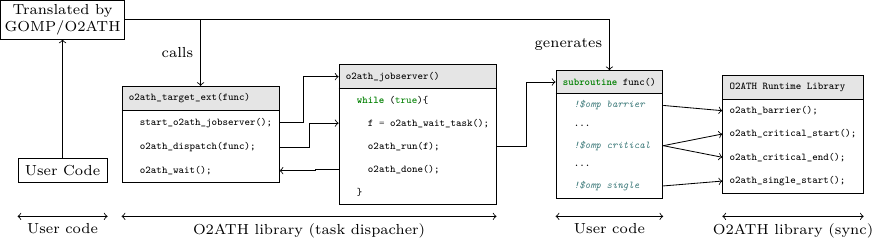}
    \caption{A detailed example for the execution workflow when using the O2ATH runtime library.}\label{fig:o2ath-run}
\end{figure*}

The most important functionality of O2ATH runtime library is to support task dispatching. As shown in Figure \ref{fig:o2ath-run}, O2ATH replaces the runtime library calls by GOMP with our O2ATH library, thus enabling the task dispatching to CPEs.

\subsubsection{Reducing Task Launch Latency}
To reduce the task launching latency, our O2ATH library spawns job servers on all CPEs with Athread library. The job server keeps alive until another task spawned by \texttt{athread\_spawn} occupies the CPEs array. This is done by wrapping \texttt{athread\_spawn} function and shutting down the job server before the real \texttt{athread\_spawn}. The task server polls an LDM variable to wait for new tasks. MPE only interacts with the task server on CPE 0, and tasks are spawned by writing its entrance function pointer and arguments into CPE 0's LDM. When CPE 0 receives a task, it broadcasts task entrance and arguments to task servers on other CPEs. After the task finishes, CPEs flush their data cache and synchronize with each other. Then, CPE 0 clears the task entrance pointer in its LDM. MPE polls the task entrance pointer in CPE 0's LDM to determine whether the task is done. The task server is written in assembly code to grant the highest efficiency. Typically, the task launch latency is less than 700 clock cycles, which is even shorter than quick thread mode in Athread library.

\subsubsection{Synchronization Directives Support}
The mostly used synchronization directives in OpenMP are \texttt{master}, \texttt{single}, \texttt{barrier} and \texttt{critical}. GOMP can generate target independent code for \texttt{master} directive, which is also used for \texttt{single} directive in O2ATH. The \texttt{barrier} and \texttt{critical} directives are implemented with CPE's \texttt{synr/sync} instruction and atomic lock, respectively.

\subsubsection{Switchable CPE Stack}\label{subsec3}
Developers usually prefer to put the CPE stack in LDM for better efficiency. But GOMP puts all private variables in stack memory, so there is a risk that LDM cannot adopt the stack. Hence, we introduce two ``virtual" devices with one CPE array. The \texttt{device 0} uses LDM as the CPE stack, and \texttt{device 1} uses CPE's private memory as the stack. O2ATH job sever can set stack pointer according to the device number.

\section{Usage of O2ATH}\label{sec4}
\begin{figure*}[ht]
  \centering
  \begin{subfigure}[b]{1\linewidth}
    \centering
    \begin{minipage}[t]{\linewidth}
      \begin{lstlisting}[style=FortranStyle]
subroutine exam(B)
!$omp declare target
  ...
end subroutine exam
...
!$omp target private(tid) device(0)
call get_coreid(tid)
DO j=js+tid, je, 64
    DO i=1, 10
        call exam(B(i,j))
        A(i,j)=A(i,j) - B(i,j)
    END DO
END DO
!$omp end target
...
      \end{lstlisting}
      \caption{The example code for accelerating the parallelizable loop.}
      \label{example-general}
    \end{minipage}
  \end{subfigure}%
  \hfill
   \begin{subfigure}[b]{1\linewidth}
    \centering
    \begin{minipage}[t]{\linewidth}
      \begin{lstlisting}[style=FortranStyle]
...
!$omp target private(gid, tid, gcnt, tcnt, tsize) device(1)
call get_vnestid(gid, tid, gcnt, tcnt, tsize)
DO k=ks+gid, ke, gcnt
    DO j=js+tid, je, tcnt
        DO i=1, 10
            A(i,j,k)=A(i,j,k) * B(i,j,k)
        END DO
    END DO
END DO
!$omp end target
...
      \end{lstlisting}
      \caption{The example code for fine-grained accelerated parallelizable loop.}
      \label{example-group}
    \end{minipage}
  \end{subfigure}%
  \caption{Two approaches to accelerate parallelizable loops with O2ATH.}
  \label{example-code}
\end{figure*}
O2ATH is a convenient toolkit with the following usage process:
\begin{enumerate}[label=\arabic*:]
    \item Replace the compiler in a project, such as mpif90, with the script \textit{xfort.py}. 
        It can compile the source file into the object file (\texttt{.o}) as well as link objects into the ELF file with O2ATH runtime library.
    \item As shown in Figure \ref{example-general}, write code for parallelizable loops as follows:
          \begin{itemize}
          \item Use the \texttt{!\$omp target} directive to indicate the start of the parallel region.
          \item Declare private variables with \texttt{private} clause in the \texttt{!\$omp target} directive.
          \item Assign a value to each private \texttt{tid}, where \texttt{tid} represents the CPE number. 
          \item Next, distribute different levels of loops for each CPE. 
          \item Finally, use the \texttt{!\$omp end target} directive to indicate the end of the parallel region. 
          \item Additionally, if the parallel region includes function calls, such as the \textit{exam} function, users need to add a \texttt{!\$omp declare target} declaration directive inside the \textit{exam} function's definition.
          \end{itemize}
          \item User has the freedom to call other CPE functions in parallelized region, as shown in Fig~.\ref{example-group}: with \textit{libvnest} library developed for CESM, K-level loops is distributed to each group and j-level loops to each CPE within the group. 
\end{enumerate}

\begin{figure*}
\centering
\includegraphics[scale=0.3]{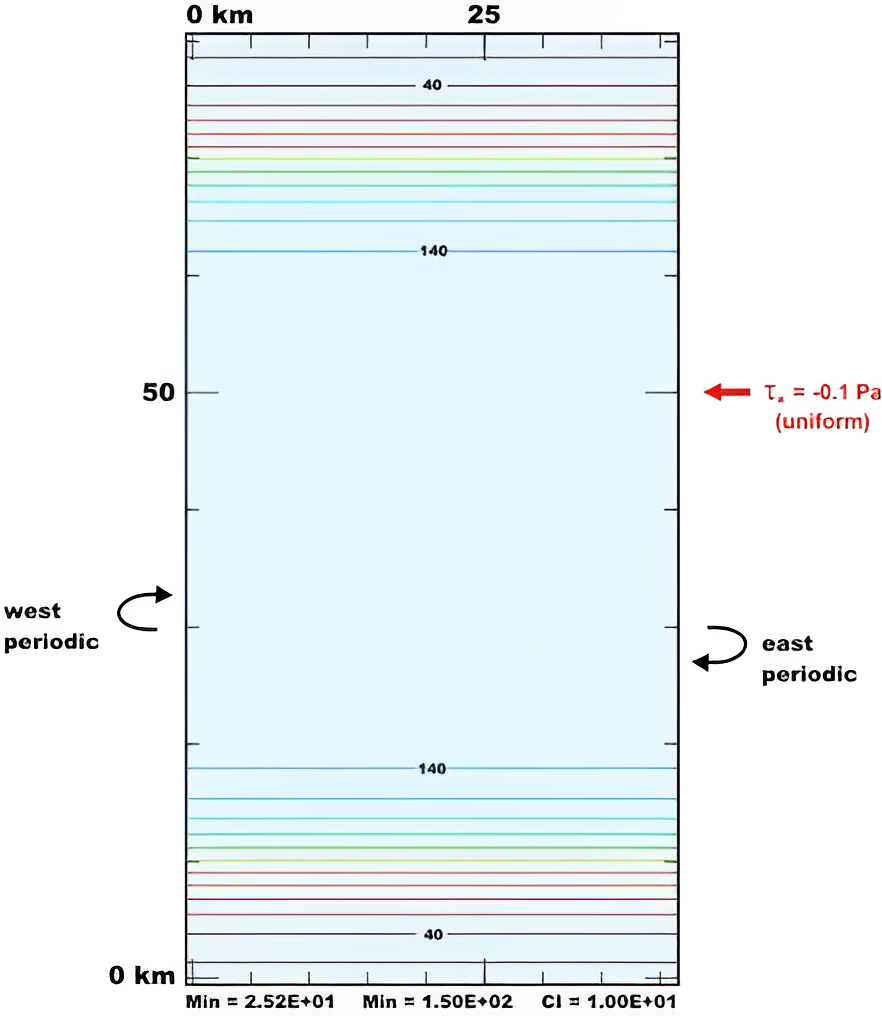}
\caption{ROMS-case1:Idealized Environment Simulation.}
\label{fig::case1_pic}
\end{figure*}
\section{Results}\label{sec5}

\subsection{Overview}\label{subsec5}
To evaluate the performance of O2ATH, we test it in two large-scale Fortran projects on the next generation Sunway supercomputer, namely CESM (2.2) and ROMS (4.0).

CESM (Community Earth System Model) is a comprehensive scientific computational model for simulating the Earth's climate system.
It finds widespread application in exploring various components of the Earth system, including the atmosphere, oceans, land, and sea ice, along with their intricate interactions.
CESM aims to facilitate a deeper comprehension of climate dynamics and the prediction of climate change and its associated impacts.
In CESM, CAM (Community Atmosphere Model) \cite{li2021enabling} is an atmospheric model designed for simulating the Earth's atmospheric system, and it serves as the largest component of CESM. We take CAM as an evaluation environment of O2ATH.

ROMS (Regional Ocean Model System) is a regional ocean model extensively employed for simulating motions across various scales, ranging from large-scale global circulation to the movement of water bodies in small-scale features such as river channels.
It finds application in diverse domains, including ocean-atmosphere coupling, marine biology, oceanic geology, and sea ice studies.
As the fields of geophysics and climatology continue to evolve, the ROMS model is increasingly utilized for high-resolution, large-scale numerical experiments involving simulations of small- to medium-scale motions.
Consequently, there is a growing demand for optimizing the performance of the ROMS model on high-performance computing platforms.

\subsection{Performance of CAM in CESM}\label{subsec5}

\begin{table*}[]
\centering
\caption{O2ATH optimizes CAM ATN\_RUN effects.}
\label{tab::cesm_performance}
\begin{threeparttable}
\begin{tabular}{ccrrr}
\toprule
Component                   & Kernel Name        & \multicolumn{1}{c}{Init Time(s)} & \multicolumn{1}{c}{Opt Time(s)} & \multicolumn{1}{c}{Speed Up\tnote{*}} \\
\midrule
\multirow{7}{*}{DYN\_RUN}   & euler\_step                       & 1.9 	                & 0.27                      & 6.9                          \\
                            & compute\_and\_apply\_rhs          & 3.2 	                & 0.82                      & 3.9                        \\
                            & advance\_hypervis\_dp             & 3.6 	                & 0.76                      & 4.7                          \\
                            & compute\_omega                    & 1.6 	                & 0.2                      & 8.0                          \\
                            & run\_consistent\_se\_cslam        & 4.3 	                & 0.7                      & 6.1                         \\
                            & tensor\_lagrange\_interp          & 1.1 	                & 0.1                      & 11.0                        \\
                            & DYN\_RUN total                    & 13.1 	                & 3.9                      & 3.4                         \\
\midrule
\multirow{3}{*}{PHYS\_RUN}  & phys\_run1                        & 8.0	                & 1.2                      & 6.7                       \\
                            & phys\_run2                        & 2.5 	                & 0.2                      & 12.5                          \\
                            & PHYS\_RUN total                   & 10.5 	                & 1.4                      & 7.5                         \\
\midrule
\multirow{1}{*}{ATM\_RUN}  & ATM\_RUN total                     & 24.2 	                & 5.5                      & 4.4                       \\
\bottomrule
\end{tabular}
\begin{tablenotes}
\footnotesize
    \item[*] The speedup does not take into account the communication time within the functions, only the computation time within the functions is considered.
\end{tablenotes}
\end{threeparttable}
\end{table*}
We use the ne30 grid configuration with 1800 processes.
Table \ref{tab::cesm_performance} compares the runtime of simulating 150 hours for various kernels before and after O2ATH optimization. 
We can see that the entire ATM\_RUN experienced a speedup of 4.4 times, with individual kernels ranging from 3.2 to 12 times improvement. 
Notably, the acceleration effect for computationally intensive physical scheme (PHYS\_RUN) was approximately 7.5 times.

\begin{table*}[]
\centering
\caption{Performance comparison of CAM DYN\_RUN kernel using O2ATH and manual optimization.}
\label{tab::cesm_per_comp}
\begin{threeparttable}
\let\cline\cmidrule
\begin{tabular}{c@{\extracolsep{8ex}}rr@{\extracolsep{8ex}}rr}
\toprule
\multirow{2}{*}{Kernel Name}  & \multicolumn{2}{c}{Speed Up}  & \multicolumn{2}{c}{Lines of Code\tnote{b}} \\\cline{2-3}\cline{4-5}
& \multicolumn{1}{c}{O2ATH} & \multicolumn{1}{c}{Manual\tnote{a}} & \multicolumn{1}{c}{O2ATH} & \multicolumn{1}{c}{Manual\tnote{a}}\\
\midrule
euler\_step                 & 3.2                               & 5.5                               & 35               &  1100\\
compute\_and\_apply\_rhs    & 3.2                               & 2.6                               & 30               &  800\\
advance\_hypervis\_dp       & 3.3                               & 1.6                               & 25               &  200\\
\bottomrule
\end{tabular}
\begin{tablenotes}
\footnotesize
\item[a] Manual means handwriting brand new slave functions to use CPEs to speed up the program.
\item[b] Code Line indicates the lines of codes to be modified compared to the initial code.
\end{tablenotes}
\end{threeparttable}
\end{table*}

Furthermore, to contrast the impact of O2ATH optimization against manual efforts, we compared the acceleration results of the dynamical framework (DYN\_RUN) with a prior study that solely utilized manually crafted CPE functions for accelerating CESM \cite{zhang2020optimizing}.
The ``Manual'' column in Table \ref{tab::cesm_per_comp} shows the speedup of the handwritten slave function, while the ``O2ATH'' column shows the speedup using O2ATH.
The ``Code Line'' column indicates the lines of code requiring modification compared to the original code.
We can see that the performance when using O2ATH is comparable with manual coding, but the reduction in the lines of code requiring modification significantly enhances development efficiency.


\subsection{Performance on ROMS}\label{subsec5}

\begin{table*}[]
\centering
\caption{Introduction to cases used in ROMS testing.}
\label{tab::case_info}
\begin{threeparttable}
\begin{tabular}{ccccccc}
\toprule
Case        & Domain\tnote{a}       &  Grid\tnote{b}    & \begin{tabular}[c]{@{}c@{}}Vertical \\ Layer\end{tabular}      & \begin{tabular}[c]{@{}c@{}}Horizontal\\ Resolution\tnote{c}\end{tabular}    & Time Step\tnote{d}  & \begin{tabular}[c]{@{}c@{}}Number of\\ Processes\end{tabular}   \\
\midrule
Case1       & \begin{tabular}[c]{@{}c@{}}145E-185E\\ 25N-50N\end{tabular}     &  135*108          & 50                & 30km                              & 360s         & 6 \\
Case2       & \begin{tabular}[c]{@{}c@{}}99E-270E\\ 8N-66N\end{tabular},       &  1522*739         & 50                & 9km                               & 150s        & 144 \\
\bottomrule
\end{tabular}
\begin{tablenotes}
\footnotesize
\item[a] The extent of the grid.
\item[b] The grid partitioning.
\item[c] The side length of each grid cell.
\item[d] The duration in physical seconds for each computational step.
\end{tablenotes}
\end{threeparttable}
\end{table*}

We conducted performance evaluations of O2ATH using two distinct test cases.
Table \ref{tab::case_info} shows the specific information of the two cases.

\begin{figure*}
\centering
\includegraphics[scale=0.2]{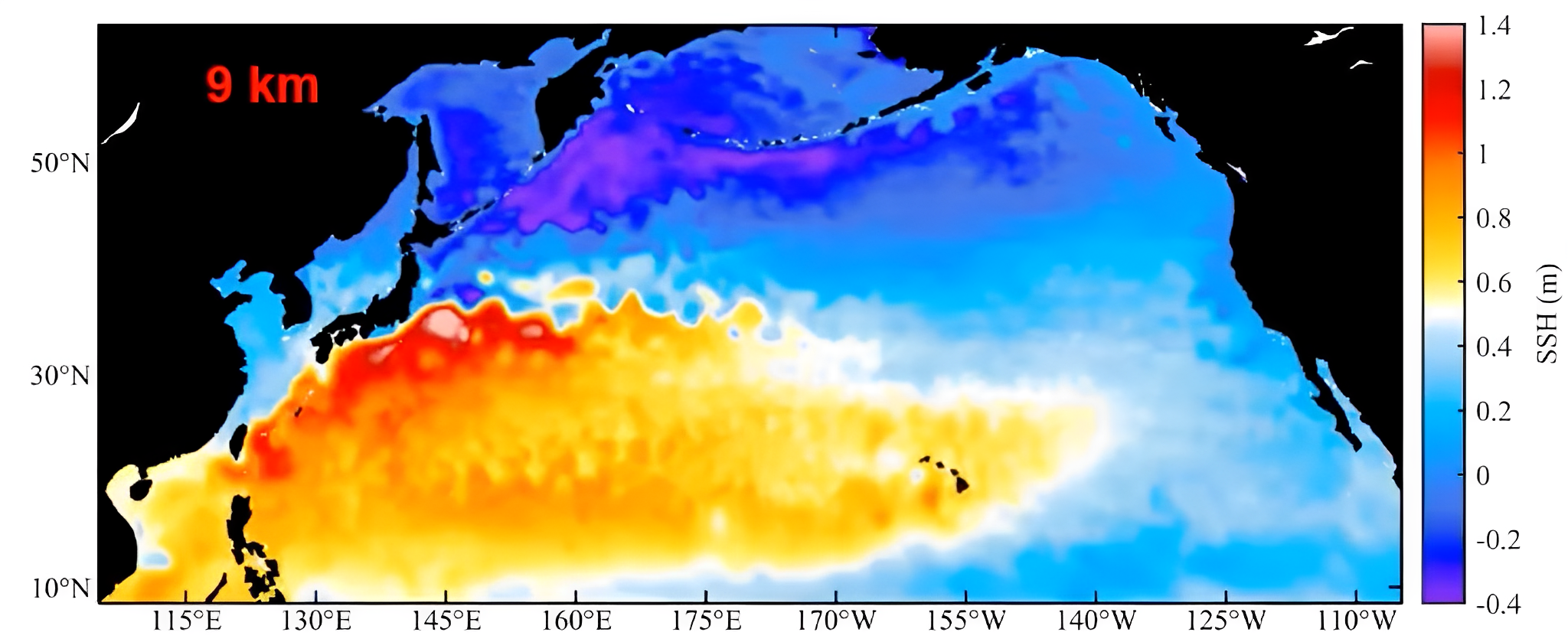}
\caption{ROMS-case2:Realistic Environment Simulation.}
\label{fig::case2_pic}
\end{figure*}

Case 1 involves an idealized scenario of a north-south temperature section, with higher temperatures at lower latitudes and cooler temperatures at higher latitudes.
There is a wind $\tau_x$ blowing throughout the region, as depicted in Figure \ref{fig::case1_pic}.
The boundary conditions for Case 1 are set as periodic, isolating the evolution of an idealized temperature field within a closed domain, thereby excluding external influences.
Conversely, Case 2 represents a real-world scenario encompassing geographical features, variations in temperature and salinity, and flow patterns, as illustrated in Figure \ref{fig::case2_pic}.
In contrast to Case 1, Case 2 entails more intricate algorithms to account for horizontal and vertical temperature gradients and flow velocity gradients.
It incorporates genuine discrete methods and vertical profiles, rendering it a more complex and computationally demanding simulation.

\begin{table*}[]
\centering
\caption{O2ATH optimizes ROMS effects.}
\label{tab::roms_performance}
\begin{threeparttable}
\begin{tabular}{ccrrr}
\toprule
Case                   & Kernel Name        & \multicolumn{1}{c}{Init Time(s)} & \multicolumn{1}{c}{Opt Time(s)} & \multicolumn{1}{c}{Speed Up} \\
\midrule
\multirow{5}{*}{Case1} & t3dmix4       & 28.2                          & 4.8                          & 5.9                          \\
                       & mpdata\_adiff & 54.2                          & 4.1                          & 13.2                         \\
                       & rhs\_3d       & 11.9                          & 1.2                          & 9.9                          \\
                       & lmd\_vmix     & 35.1                          & 7.0                          & 5.0                          \\
                       & prsgrd        & 6.6                           & 0.6                          & 11.0                         \\
                       & total        & 341.0                           & 64.5                          & 5.3                         \\
\midrule
\multirow{5}{*}{Case2} & t3dmix4       & 174.0                           & 15.7                         & 11.1                         \\
                       & mpdata\_adiff & 485.0                         & 55.6                         & 8.7                          \\
                       & rhs\_3d       & 62.8                          & 6.1                          & 10.3                         \\
                       & lmd\_vmix     & 220.4                         & 34.1                         & 6.5                          \\
                       & prsgrd        & 39.0                          & 3.1                          & 12.6         \\               
                       & total        & 1200.0                          & 340.0                          & 3.5         \\               
\bottomrule
\end{tabular}
\end{threeparttable}
\end{table*}

Table \ref{tab::roms_performance} presents the acceleration outcomes achieved before and after applying O2ATH to two distinct cases.
It can be seen that the computational kernels utilizing O2ATH have achieved acceleration factors of 5 to 13.
In addition, the overall performance of the software has been improved by approximately 3 to 5 times.

\begin{table*}[]
\centering
\caption{Performance comparison of partial ROMS lmd\_bkpp kernel using O2ATH and manual optimization.}
\label{tab::roms_per_comp}
\begin{threeparttable}
\begin{tabular}{crrrr}
\toprule
Case  & \multicolumn{1}{c}{O2ATH Time(s)} & \multicolumn{1}{c}{Manual Time(s)\tnote{a}} & \multicolumn{1}{c}{O2ATH Code Line\tnote{b}} & \multicolumn{1}{c}{Manual Code Line} \\

\midrule
Case1 & 5.7                               & 4.8                                & 4               & \textgreater{}100 \\
Case2 & 17.1                              & 21.2                               & 4               & \textgreater{}100 \\
\bottomrule
\end{tabular}
\begin{tablenotes}
\footnotesize
\item[a] Manual means handwriting brand new slave functions to use CPEs to speed up the program.
\item[b] Code Line indicates the lines of codes to be modified compared to the initial code.
\end{tablenotes}
\end{threeparttable}
\end{table*}

In addition, we also selected a portion of the lmd\_bkpp kernel for manual optimization.
Table \ref{tab::roms_per_comp} shows the analysis of O2ATH compared to the manual optimization, including the performance comparison and the number of modified lines of code.
We can see that O2ATH has about 15-20\% performance loss compared to manual optimization.
However, it is worth noting that O2ATH modifies significantly fewer lines of code compared to the manual optimization process.
This means that although O2ATH compromises slightly on performance, it makes far fewer changes to the original code base.
Therefore, O2ATH is an efficient solution for software development, especially when it comes to porting or optimizing large software systems.

\section{Future Work}\label{sec6}
In seek of low task latency, O2ATH currently cannot support nested parallelism in a standard OpenMP way, and this will be improved during further development of O2ATH. We plan to support the two-level parallel scheme in OpenMP, namely teams-threads, to replace \texttt{libvnest} used in CAM. Also, the parallel loop is not well-supported because of the missing of O2ATH library routines, and we plan to support static and dynamic scheduling with user configurable tiling size.
MPE and CPE do not share the same vector extensions, and this is not solved in O2ATH library. Maybe a middleware can be developed to yield a unified vector interface.
On the virtual function and trampolines support on CPEs, we also plan to do further plugin work to make them supported in a more clean and robust way.
\section{Conclusion}\label{sec7}
In this paper, we present a proxy toolkit named O2ATH.
It includes a compiler plugin and a runtime library.
The plugin assesses whether each function should be compiled for CPE or MPE and takes steps to remove unnecessary functions from the compiler's IR. 
The runtime library supports task dispatching and some fundamental OpenMP directives. 
O2ATH offers a straightforward and convenient usage, greatly simplifying the process of migrating Fortran applications to the Sunway supercomputer. 
We have successfully accelerated two large fortran projects, CESM and ROMS, on the next generation Sunway supercomputers using O2ATH.
The experimental results demonstrate that there are speedups ranging from 3 to 15 times across different kernels with O2ATH, resulting in an overall performance improvement of 3 to 6 times for the projects.
What's more, using O2ATH required significantly fewer code modifications compared to manually crafting CPE functions, which means that using O2ATH makes it easier to maintain large-scale applications.
In practical applications, O2ATH not only saves a significant amount of time but also achieves excellent performance. 

\section{A Conflict of Interest Statement}
On behalf of all authors, the corresponding author states that there is no conflict of interest.

\end{document}